\documentclass[epj]{svjour}
\usepackage{amsmath}
\usepackage{epsfig}
\begin{document}
\title{Dielectric resonances in disordered media}
\author{Laurent Raymond, Jean-Marie Laugier, Steffen Sch\"afer and Gilbert Albinet}
\authorrunning{Raymond and al.}
\titlerunning{Resonances in disordered media}
\institute{Laboratoire Mat\'eriaux et Micro\'electronique de
Provence (UMR CNRS 6137) et Universit\'e Aix-Marseille I \\
B\^at.~IRPHE, Technopole de Ch\^ateau-Gombert, 49 rue Joliot Curie,
B.P.~146, 13384 Marseille Cedex 13, France.} 
\offprints{Laurent Raymond \email{laurent.raymond@l2mp.fr}} 
\mail{same address}
\date{August 16, 2002}

\abstract{Binary disordered systems are usually obtained by mixing two
ingredients in variable proportions: conductor and insulator, or
conductor and super-conductor. They present very specific properties,
in particular the second-order percolation phase transition, with its
fractal geometry and the multi-fractal properties of the current
moments.  These systems are naturally modeled by regular
bi-dimensional or tri-dimensional lattices, on which sites or bonds
are chosen randomly with given probabilities.  The two significant
parameters are the ratio $h=\sigma_{1}/\sigma_{0}$ of the complex
conductances, $\sigma_{0}$ and $\sigma_{1}$, of the two components,
and their relative abundances $p$ (or, respectively, $1-p$).  In this
article, we calculate the impedance of the composite by two
independent methods: the so-called spectral method, which diagonalises
Kirchhoff's Laws via a Green function formalism, and the Exact
Numerical Renormalization method (ENR).  These methods are applied to
mixtures of resistors and capacitors (R-C systems), simulating
e.g. ionic conductor-insulator systems, and to composites consituted
of resistive inductances and capacitors (LR-C systems), representing
metal inclusions in a dielectric bulk.  The frequency dependent
impedances of the latter composites present very intricate structures
in the vicinity of the percolation threshold.  In this paper, we
analyse the LR-C behavior of compounds formed by the inclusion of
small conducting clusters (``$n$-legged animals'') in a dielectric
medium. We investigate in particular their absorption spectra who
present a pattern of sharp lines at very specific frequencies of the
incident electromagnetic field, the goal being to identify the
signature of each animal. This enables us to make suggestions of how
to build compounds with specific absorption or transmission properties
in a given frequency domain.
}
\maketitle
 
\section{Introduction}
Composite materials, obtained for instance by mixing powders, are
increasingly used in modern mechanical, electrical and optical
devices.
Their extraordinary properties often meet the very compelling
standards needed for high technology materials, e.g. high temperature
resistance, low density, and low thermal or electrical conductivity.
Their manufacturing is however extremely delicate and requires a
good comprehension of the microscopic and macroscopic properties
of the different constituents.
The goal of this paper is to provide a better theoretical
understanding of various electrical properties of these materials,
especially in the high frequency domain.

Composite systems are commonly thought of as random networks where
each bond represents, via a complex impedance, a grain or a grain
boundary.
The various constituents of the composite material are then
randomly distributed over the network, and elements of the same
type connected to each other form ``clusters'', or ``animals''
in the terminology of Pierre-Gilles de Gennes.
The interfaces between clusters are usually the physically most
interesting regions \cite{Debi,Gilbert}.
Therefore, in binary composites, the percolation of one component
through the other plays an essential role: the properties of the
material change dramatically for small variations of the chemical
composition in the vicinity of the percolation threshold, giving rise
to a second-order phase transition which allows the physicist to put a
large number of disordered systems in the same universality class.

These heterogeneous media occur mainly as bulk material occupying
a 3D volume, or as thin, almost 2 dimensional, layers upon a
substrate.
In both cases, the electrical properties, {\sl i.e.} the
frequency-dependent network impedance, can be obtained as a direct
solution of Kirchhoff's Laws for each network node. As this involves
the diagonalisation of large matrices, this method becomes very
expensive in CPU time as soon as realistic systems are to be modelled.
On the other hand, a crude mean-field approximation, although
generally qualitatively correct, is not sophisticated enough to
reproduce experimental results quantitatively.

Alternative approaches are provided by spectral methods, based on the
theory of random walks (see the excellent article by W.H.~McCrea and
F.J. Whipple \cite{McCrea}, and the book of F.~Spitzer
\cite{Spitz}). These approaches have been extended to percolation
phenomena \cite{Clerc2,Luck} and work particularly well in two
dimensions. For a random 2D network, the electrical and optical
properties are described by its analytically obtainable conductance
poles (resonances) and their residues (weights). This approach can
also be extended to binary disordered media in 3D, as shown by two of
the authors in a previous paper, although the possibility of an
analytic treatment is lost \cite{Laurent}.

One goal of the present paper is to study the elementary clusters,
or ``animals'', which take part in percolation until the threshold
is reached and the cluster becomes infinite. A detailed
description of an algorithm generating all animals for a given
number of bonds (``legs''), and the complete ``zoo'' of up to
4-legged animals will be given in the appendix.
A composite's response to a frequency-dependent electromagnetic signal
is a highly characteristic spectrum which can be viewed as the
signature of conducting clusters (animals) embedded in a dielectric
medium.  
In the dilute limit, where the influence of one animal on its
neighbors is negligible, one measures the almost unperturbed spectral
fingerprint of the individual animals.  In this case, a theoretical
study of the spectra of a limited number of small, elementary animals
provides a good starting point for the interpretation of the response
of the composite as a whole.  Still provided the concentration of
metallic clusters is low, one may even be confident to gain some
insight into the microscopic structure of the material itself.

Our aim is to compare these results to those of the renormalization
algorithm described below. The latter remains very efficient even if
the animals become large, or if elementary animals are arranged as
regular arrays over the whole lattice, cases for which the exact
spectral calculations become very cumbersome. The algorithm
implemented for performing these simulations is called Exact Numerical
Renormalization (ENR) and was initially proposed in
Ref.~\cite{Vinograd,Sarych,Tort}. The basic idea is to eliminate
network nodes successively, and to connect all neighbors of the
eliminated site by bonds with renormalized impedances.  The method is
essentially applicable to any connected network, regardless of
dimensionality and connectivity. In the following, however, we will
restrict our considerations to the hypercubic lattices in two and
three dimensions. We will evaluate the impedance versus frequency
curves of metallic animals, constituted of bonds, or legs, to which a
complex conductivity is attributed, on square (2D) and simple cubic
(3D) dielectric lattices.

The notations of this paper are those of Ref.~\cite{Clerc2,Clerc1}; we
recall them shortly. The bond occupation is obtained according to the
following binary law: conductance $\sigma_{0}$ and concentration $p$
for one kind of bond (representing the animal's legs), $\sigma_{1}$
and $1-p$ for the rest of the lattice (voids). The dimensionless
complex ratio $h=\sigma_{1}/\sigma_{0}$ of the two conductances and
the relative abundance $p$ are the essential parameters of the model.
For convenience, $h$ may be replaced by the equivalent complex
variable
\begin{equation}
\label{lambda}
\lambda =  \frac{1}{1-h} =
\frac{\sigma_{0}} {\sigma_{0} - \sigma_{1}}
\end{equation}

For the 2D case, the square lattice is self-dual, and quantities such
as the percolation threshold, or spectral properties of some animals,
can be easily deduced from this particularity. In 3D, the self-duality
property is lost.  In section \ref{sect1}, we recall the analytic
method and the numerical algorithm. In section \ref{sect2}, we examine
the spectral properties of elementary animals and the particularities
which arise if such metallic animals are disposed as regular
super-arrays on a dielectric lattice; the real and imaginary parts of
the impedance are presented and discussed for a large panel of such
animals. In section~\ref{sect3}, we are concerned by binary random
lattices and the corresponding Nyquist diagrams. The recursive
algorithm used for the creation of all $n$-animals (with $n$ being the
number of legs), and the symmetry properties of animals with $n\le 9$
are discussed in the appendix.  We conclude the paper by proposing
some applications of binary random networks.

\section{Model and algorithm}
\label{sect1}

We consider a binary composite constituted of a random distribution of
electrical bonds and voids on a hypercubic lattice. The total
conductance $Y$ of the sample (or, alternatively, its impedance
$Z={1}/{Y}$), is obtained by the fulfillment of Kirchhoff's Current
Law at each network node $x$:
\begin{equation}
\sum_{y(x)}\sigma_{x,y}(V_{x}-V_{y}) = \sum_{y(x)} I_{x,y} = I_x
\end{equation}
with $V_{x}$ corresponding to the potential at node $x$, $I_x$ the
current arriving on the node $x$, and $I_{x,y}$ the current from $y$
to $x$ along the link of conductance $\sigma_{x,y}$.
If a total current $I$ flows through the sample between the two
electrodes, of which one is at potential $V_{1}$ and the other
grounded ($V=0$), the conductance of the network reads $Y=I/V_{1}$.
Only the finite section ${\cal L}$ of a square (2D) or cubic lattice
(3D) in-between the two electrodes is considered. In three dimensions,
this piece is characterized by $$ {\cal L} = \{1 \dots N_x \} \times
\{ 1 \dots N_y \} \times \{ 1 \dots (N_z-1) \} \, .  $$ The plans
$z=0$, and $z=N_z$ (respectively $y=0$, and $y=N_y$ in the 2D case)
are taken as electrodes. Each link of this binary network is assumed
to have either conductance $\sigma_0$ with probability $p$, or
conductance $\sigma_1$ with probability $q=1-p$.
In order to reduce finite-size effects, periodic boundary conditions
are imposed in all directions parallel to the electrodes, {\sl
i.e.} in the $x$ ($x$ and $y$) direction in 2D (3D, respectively).

A convenient alternative to a direct solution of Kirchhoff's Laws is
the spectral method in which the spectrum results from a solution of a
generalized eigenvalue problem. This spectrum presents a rich set of
resonances, characteristic of the bond distribution and the underlying
lattice structure.  This approach, proposed by Straley \cite{Stra} and
Bergman \cite{Berg}, yields the conductivities, corresponding to
different values of $h=\sigma_{1}/\sigma_{0}$, by means of the
frequencies and weights of the conductance poles.  Instead of the bare
$h$ itself, one may of course use the handier parameter
$\lambda={1}/(1-h)$ defined in eq.~(\ref{lambda}) which confines the
poles to the interval $[0,1]$.

It has been adapted to 2D finite networks in Ref.~\cite{Luck} and
enhanced to 3D in Ref.~\cite{Laurent}.  However, as larger networks
are to be treated, the method becomes very time-consuming, since it
involves the numerical diagonalisation of large matrices. In the
following, we will therefore tackle the problem with another algorithm
which is inspired from the renormalization procedure
\cite{Vinograd,Sarych,Tort}.

In this method, known as Exact Numerical Renormalisation (ENR), the
network sites are eliminated one by one. At each step, the former
neighbors of the eliminated site are connected by possibly new bonds
whose impedances are chosen such that the global impedance of the
system remains invariant.
Namely, if $x$ is the site to be eliminated, the conductivity
$\sigma_{i,j}$ between all sites $i$ and $j$ of the neighborhood of
$x$ will be reassigned to $\sigma_{i,j}^{*}$. It is easy to verify
that $\sigma_{i,j}^{*}$ is given by
\begin{equation}
\sigma_{i,j}^{*} = \sigma_{i,j} +
\frac{\sigma_{i,x}\sigma_{x,j}}{\sum_{j}{\sigma_{x,j}}} \, .
\end{equation}
where the summation is over all neighbors of site $x$ \cite{Tort}. 
Note that $\sigma_{i,j}=0$ if sites $i$ and $j$ are not connected
before the elimination of $x$. The renormalization procedure is
numerically exact. The ENR procedure stops when the two electrodes are
linked by only one bond which then carries the total conductivity of
the initial lattice. Details of calculation are given in
Ref.~\cite{Tort}.  The ENR method is particularly well adapted to a
sparse medium, and can be applied to any linear network including
systems with more than two components.  For instance in a
conductor-insulator system near the percolation threshold, each site
$x$ of the infinite cluster has an average number of neighbors smaller
than the connectivity of the system, and a great number of sites can
be suppressed without the introduction of additional links between
remaining sites.  But even for the problems considered in this paper,
where all bonds of the lattice must be taken into account and the
renormalization has to be performed for all frequencies, the ENR
algorithm is considerably faster than the spectral method developed in
Ref.~\cite{Luck,Laurent}. The main restriction of the ENR method is
that the nature of the components has to be determined by defining a
relation $h(\omega)$ or $\lambda(\omega)$ {\em before} the
computation, e.g. $h(\omega)={\rm j}C\omega (R+{\rm j}L\omega)$ for
the LR-C model considered in the following.

This paper is devoted to the study of:
\begin{enumerate} 
  \item {Small clusters (``animals'') of a few connected conducting
  bonds (of inductance L and resistance R) in an insulating
  environment (of capacity C) \cite{Clerc2}, the aim being to detect
  the signature of elementary clusters in the spectral response of a
  conductor--insulator mixture.}

  \item {Percolation through a binary Resistive-Capacitive
  lattice. For these systems, one has to sample over a large number of
  random networks in order to gain some information which is intrinsic
  to the chemical composition $p$.  For disordered systems, one
  specific random network is not interesting, and all quantities have
  to be averaged over a sufficiently large number of systems of the
  same degree of disorder. For obvious symmetry considerations, the
  interchange of $p$ and $q$ corresponds to replacing $h$ by
  ${1}/{h}$, or $\lambda$ by $1-\lambda$. This symmetry relation
  allows us to confine the range of investigation of such models to
  $p\le\frac{1}{2}$. Moreover, one can immediately conclude that any
  averaged quantity computed for $p=\frac{1}{2}$ will be symmetrical
  around the value $\lambda=\frac{1}{2}$. This implies, for instance,
  that the average conductance of the network obeys
\begin{equation}
Y(p,\sigma_0,\sigma_1) = \sigma_0 \tilde{Y}(p,\lambda)=
        \sigma_1 \tilde{Y}(1-p,1-\lambda) \, .
\end{equation}
}
\end{enumerate}

The spectral algorithm yields the positions and cross-sections of all
resonances and for any cluster (provided the cluster is not too large
and the numerical calculation remains feasible \cite{Laurent}), and a
fortiori for small $n$-animals ($n\leq 10$). In 2D, moreover, all
quantities may --- at least in principle -- be evaluated analytically
on an infinite lattice \cite{Clerc2}.

The ENR algorithm, on the other hand, is particularly well adapted to
large clusters (percolation), and to the study of the coupling between
a few animals as a function of their mutual distances. Instead of the
resonance positions (eigenvalues) and the corresponding weights (which
are connected to the eigenvectors), this algorithm yields readily the
total frequency-dependent impedance of the whole system with the
desired degree of accuracy.

\section{Spectral properties of animals}
\label{sect2}

\begin{figure}
\caption{All topologically different animals (species) with one leg (one
species), two legs (two species), three legs (five species), and four
legs (16 species). The percentages indicate how many animals belong to
a given species in a zoo.}
\label{fig1}
\begin{center}
\epsfig{figure=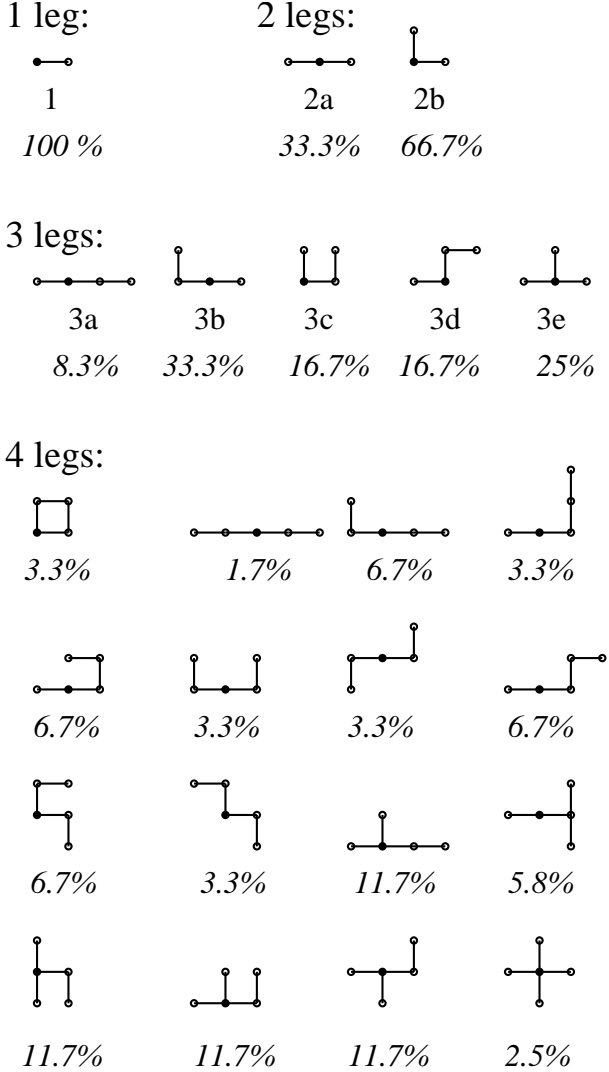,width=8cm}
\end{center}
\end{figure}

In this paragraph, small metallic clusters are investigated.  These
``animals'' are constituted of $n$ conducting bonds (``legs'' of
self-inductance $L$ and resistivity $R$), and ``live'' on an
insulating lattice of capacitors $C$ \cite{Clerc2}. For the 2D case,
an exhaustive list of all up to 4-legged animals living on a square
lattice, together with their geometries and abundances, is given in
the Fig.~\ref{fig1}.

For animals living on a finite 3D network, the spectra show pole
densities and average resonance positions which scale as the inverse
of the linear size of the sample, thus corresponding to the ratio
surface over volume for a cubic sample \cite{Laurent}. This allows one
to extrapolate the finite size results to an infinite lattice.
Alternatively, the resonance values and the pole densities of a single
animal on an infinite network can be evaluated exactly by the method
described in \cite{Clerc2}.  The positions of the resonances are then the
eigenvalues of a matrix $M_{x,y}$ obtained from the values of the
Green function of the laplacian operator on the infinite lattice
\begin{equation}
M_{x,y} = \sum_{z\in C(y)}{(G_{x,y} - G_{x,z})} \, .
\end{equation}
where $x$, $y$ and $z$ belong to the conducting cluster and the index
of summation, $z$, runs over the neighborhood of $y$, $C(y)$. In
general, this is a $n_s\times n_s$ non-symmetrical matrix, where $n_s$
is the number of sites of the animal under consideration.
\begin{table}
\caption{Exact resonance position $\lambda$ and corresponding residues
for animals consisting of one, two and three bonds. Analytical results
are compared to ENR results for a single LR animal in the center of a
$24\times 24$ square lattice. The animals' labels are those of
Fig.~\ref{fig1}.}
\label{tab1}
\begin{center}
\begin{tabular}{|c|l|l|l|l|}
\hline
animal & $\lambda$ exact & $\lambda$ ENR & residue & $Re(Z)$\\
\hline \hline
1 & 0.5 & 0.5011248 & 4. & 2.30485\\
\hline \hline
2a & 0.36338 & 0.367 & 8.6455 & 4.95521 \\
         & 0.63662 & & 0.000 &  \\
\hline
2b & 0.318310 & 0.316 & 2.304 & 1.471 \\
         & 0.681690 & 0.6784 & 2.304 & 1.09 \\
\hline \hline
3a & 0.28216 & 0.29219 & 14.633 & 9.155 \\
         & 0.54648 &  & 0.000 &  \\
     & 0.67136 & 0.67158 & 0.16406 & 0.1175 \\
\hline
3b & 0.24970592 & 0.251 & 6.959 & 4.202 \\
         & 0.54315776 & 0.5424 & 2.239 & 1.454 \\
     & 0.70713638 & 0.7061 & 0.679 & 0.4564 \\
\hline
3c & 0.16581640 &  0.167 & 2.686 & 1.8034 \\
         & 0.6366200 &  & 0.000 &  \\
     & 0.69756359 & 0.698 & 2.979 & 1.956  \\
\hline
3d & 0.22926367 & 0.243 & 2.718 & 3.693 \\
         & 0.55349146 &  & 0.000  &  \\
     & 0.55349146 & 0.7575 & 2.718 & 3.66 \\
\hline
3e      & 0.30243640 &  & 0.000 & \\
         & 0.3633800 & 0.367 & 8.645 & 5.48 \\
     & 0.8341836 &  & 0.000 &  \\
\hline
\end{tabular}
\end{center}
\end{table}
It is straightforward to obtain an animal's set of resonances (see
e.g. Table~\ref{tab1} for all animals up to 3 legs shown in
Fig.~\ref{fig1}).  The cross section ({\sl i.e.} the residue)
$\gamma_{a}$ of a resonance $\lambda_{a}$ can be obtained
analytically. For each animal, one finds a resonance at $\lambda=0$
which has no physical meaning and carries $0$ weight \cite{Laurent}.

This spectral method developed for the 2D case can be readily extended
to 3D systems, but the helpful duality property gets lost in 3D.
\begin{table}
\caption{Resonance position and corresponding residues for some one-
or two-legged animals located at various positions in a $8\times
8\times 8$ cubic lattice. Due to the boundary conditions, only the
current direction ($z$) is relevant. $i_{n}$ labels a 1-animal
consisting of a single bond between plans $z=n$ and $n+1$. $I_{n}$
represents a 2-animal consisting of two adjacent bonds along the same
direction, between plans $z=n$ and $n+2$.}
\label{tab2}
\begin{center}
\begin{tabular}{|c|l|l|}
\hline
animal & pole position & residue $10^{-2}$ unit \\
\hline \hline
$i_0$, $i_7$ & 0.789308423 & 4.17081055 \\
\hline
$i_1$, $i_6$ & 0.674277363 & 7.54795650 \\
\hline
$i_2$, $i_5$ & 0.668037619 & 7.76440136 \\
\hline
$i_3$, $i_4$ & 0.667363895 & 7.78801367 \\
\hline \hline
$I_0$, $I_6$ & 0.570490690 & 0.77324464 \\
         & 0.893095096 & 3.61773483 \\
\hline
$I_1$, $I_5$ & 0.544761964 & 0.00397830 \\
         & 0.797553018 & 7.93113971 \\
\hline
$I_2$, $I_4$ & 0.543267933 & 0.00004814 \\
         & 0.792133581 & 8.20040186 \\
\hline
$I_3$      & 0.543157473 & 0 \\
         & 0.791570318 & 8.22848890 \\
\hline
\end{tabular}
\end{center}
\end{table}
\begin{table}
\caption{Similar results as in table \ref{tab1} for a right-angle
animal with the vertex on row $n+1$.}
\label{tab3}
\begin{center}
\begin{tabular}{|c|l|l|}
\hline
animal & pole position & residue $10^{-2}$ unit \\
\hline
$\Gamma_0$ & 0.566366254 & 4.05530700 \\
         & 0.903637749 & 1.10140027 \\
\hline
$\Gamma_1$ & 0.534390544 & 6.66404562 \\
         & 0.808411116 & 1.89039118 \\
\hline
$\Gamma_2$ & 0.531898275 & 6.86342706 \\
         & 0.803701317 & 1.91149019 \\
\hline
$\Gamma_3$ & 0.531617073 & 6.88538519 \\
         & 0.803195822 & 1.91366884 \\
\hline
$\Gamma_4$ & 0.531730016 & 6.88512684 \\
         & 0.803195852 & 1.91287160 \\
\hline
$\Gamma_5$ & 0.532860589 & 6.86124029 \\
           & 0.803701327 & 1.90472208 \\
\hline
$\Gamma_6$ & 0.546552694 & 6.64049976 \\
           & 0.808420249 & 1.80603241 \\
\hline
\end{tabular}
\end{center}
\end{table}
The resonance positions given for a finite size realization of the 3D
case are presented in Table~\ref{tab2} and Table~\ref{tab3}. They show
the sensitivity of the spectra to the animals' position with respect
to the electrodes, and differ considerably from the 2D results listed
in Table~\ref{tab1}. 

In what follows, we restrict our analysis to the 2D case but a
generalization to 3D is immediate. 
\begin{figure}
\caption{Representation of a square lattice in 2D, of size
$N_x=N_y=10$. The two electrodes at $y=0$ and $y=N_y=10$ are,
respectively, at potential $V_{1}$ (assumed $>0$) and $V_{0}=0$.
Periodic boundary conditions are assumed in the $x$ direction.  Twelve
animals are shown for example.}
\label{fig2}
\begin{center}
\epsfig{figure=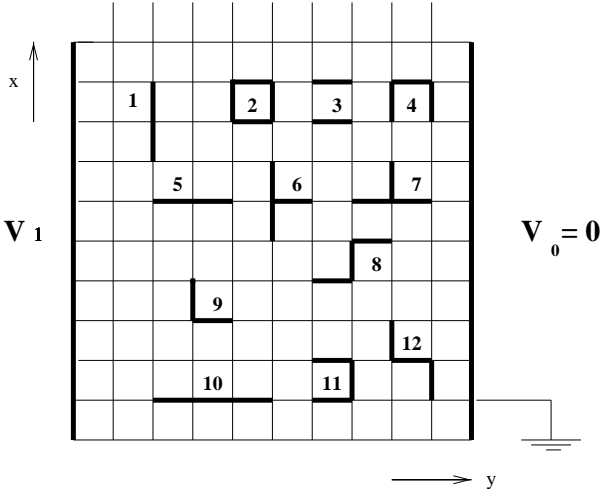,width=8cm}
\end{center}
\end{figure}
Applying the formalism proposed by Clerc {\sl et al.} \cite{Clerc2},
to the 2-legged animals 2a and 2b of Fig.~\ref{fig1}, we obtain 
$$ M =
\frac{1}{4}
\left(
\begin{array}{c c c}
1 & \, -(2+4g) \, & \, 4g+1 \\
-1 & 2 & -1 \\
4g+1 & \, -(2+4g) \, & \, 1
\end{array}\right)\, .
$$ 
The value of $g$ depends on the relative orientation of the two bonds
and may be obtained analytically in 2D: $g=G_{2}=\frac{2}{\pi}-1$ for
two bonds in a line, and $g=G_{1,1}=-\frac{1}{\pi}$ for two orthogonal
bonds. One obtains $\lambda_{1}=-g$ and $\lambda_{2}=g+1$ as
eigenvalues of $M$ leading to resonances at $h_{1}=\frac {1+g}{g}$ and
$h_{2}=\frac{g}{1+g}$. 
For self-dual animals in 2D, duality implies that the solutions occurs
as pairs verifying the relations $\lambda_{1}+\lambda_{2}=1$ or $h_{1}
h_{2}=1$.
For two adjacent links with the same orientation, we have
$g=G_{2}=-0.36338$, and the two resonances occur respectively at
$\lambda_{1}=0.36338$ and $\lambda_{2}=0.63662$. For two orthogonal
links, we have $g=G_{1,1}=-0.31831$, and the resonances are
$\lambda_{1}=0.31831$ and $\lambda_{2}=0.68169$ (see
Table~\ref{tab1}). In both cases, these values are well confirmed by
the ENR results (except that, for obvious reasons, the zero weight
resonance of animal 2a cannot be detected by ENR).

For all up to 3-legged animals represented in Fig.~\ref{fig1}, the
exact and numerical eigenfrequencies, and the corresponding exact
residues along with $Re(Z)_{\rm max}$ are listed in Tab.~\ref{tab1}.
The spectrum is deduced from the numerically obtained real part of the
impedance versus $\omega$ or $\lambda$; within the ENR algorithm the
residue of each pole is closely related to the maximum value of the
real part of the impedance, $Re(Z)_{\rm max}$ \cite{Clerc2,Luck}.
For a 2D $N\times N$ lattice with $L=C=1$ and resistance $R$ varying
between $1\cdot 10^{-3}$ and $2\cdot 10^{-1}$, we find:
\begin{equation}
 Re(Z)_{\rm max} \approx \frac{\gamma_{a}}{R N^2} 
\end{equation}
For an animal in the center of a sufficiently large lattice (e.g.
$N_{x}=N_{y}=24$) finite size effects -- resulting in a small shift of
the resonance positions -- may be safely ignored. Its spectrum may be
calculated for pulsations $\omega$ such that $\lambda(\omega)$ runs
over the complete interval $[0,1]$.

It is numerically impossible to find resonances of zero
cross-section. Therefore, animals of the same species, which
necessarily have the same eigenvalues $\lambda$ --- as e.g. in
Fig.~\ref{fig2} the animals $1$ and $5$, $6$ and $7$, $4$ and $11$,
and $8$ and $12$ --- may be oriented differently with respect to the
electrodes in order to identify, if possible, the positions of the
resonances of zero weight. According to this analysis, the animal $1$
of Fig.~\ref{fig2}, whose three sites are on the same potential, is
not crossed by any current: consequently, the three resonances have
zero weight and do not contribute to the conduction.  The poles
corresponding to diagram $7$ of Fig.~\ref{fig2} --- a ``T'' with the
two bonds of its horizontal bar perpendicular to the electrodes ---
are two times stronger than those of diagram $6$ --- the same ``T''
but turned by $90^\circ$ such that only one bond is perpendicular to
the electrodes.  The same holds true, respectively, for the diagrams
$8$ and $12$, and also for $4$ and $11$.  The diagrams $2$, $3$ and
$11$ present strictly the same response to an incident a.c. wave:
obviously, the bonds parallel to the electrodes do not participate to
the resonance phenomena at all.

For large clusters, or when several clusters are present on the
surface, it is very lengthy to obtain the exact spectral responses
since inter-cluster interactions must be taken into account.  Under
these conditions, the ENR algorithm reveals its efficiency. We only
present the calculations for the 2D case. In 3D, the physical ideas
are essentially the same, but the calculus becomes very cumbersome.
\begin{figure}
\caption{Real part $Re(Z)$ of the impedances versus $\lambda$ for
various animals living on a $12\times12$ square lattice.
(a) The continuous peak at $\lambda=0.5$ is the signature of the
simplest 1-animal --- a single inductance --- in $y$-direction. The
dashed peak on its right, at $\lambda\simeq 0.63$, is obtained with
two parallel coils on successive lines (animal no.~3 in
Fig.~\ref{fig2}), and the dotted-dashed peak on the left, at
$\lambda\simeq 0.37$, has its origin in two adjacent inductances on
the same line (animal no.~5 in Fig.~\ref{fig2}).
(b) The dashed peak on the right was obtained by two parallel coils in
$y$-direction, separated by an empty line. The dot-dashed peak on the
left corresponds to two coils on the same horizontal line, separated
by a capacitive link. Comparison with the dashed and dot-dashed curves
in Fig.~\ref{fig3}a shows that the coupling between the coils
decreases and the resonances occur closer to the value $\lambda \simeq
0.50$ (characteristic for a single coil).
(c) Spectra given by various staircase-shaped animals: the continuous
line, with the two peaks closest to the center, at $\lambda=0.32$ and
$\lambda=0.68$, origins from two coils forming a right-angle (animal
no.~9 in Fig.~\ref{fig2}).  The dashed line, showing peaks at
$\lambda\simeq 0.22$ and $\lambda\simeq 0.78$, corresponds to animals
no.~8 and 12 in Fig.~\ref{fig2}.  The dot-dashed line with the
outermost peaks, at $\lambda\simeq 0.16$ and $\lambda\simeq 0.84$
corresponds to a four-step staircase animal.  The separation in
$\lambda$ between an animal's two spectral peaks thus increases with
the animal's size.
(d) Comparison of two disordered system with the same densities of
coils: the full line corresponds to 36 two-bond right angles, with
vertex sites chosen at random. The dashed line corresponds to 72
randomly chosen bonds, 36 of which are horizontal and the other 36
vertical.} 
\label{fig3}
\begin{center}
\epsfig{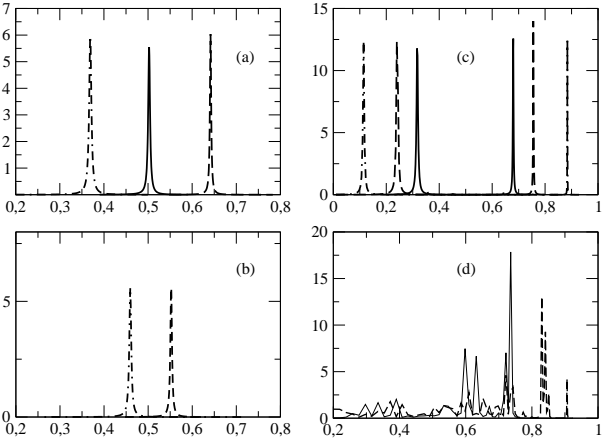}
\end{center}
\end{figure}
Fig.~\ref{fig3}, where the weights of the resonances are represented
as a function of the parameter $\lambda$, shows the coupling between
resistive coils placed in different locations of the capacitive
lattice. In the spectra represented on Fig.~\ref{fig3}a, each
inductance ($L$,$R$) occupies a link in the $y$-direction
(perpendicular to the electrodes) and it is not connected to any
electrode.  The central peak corresponds to two well separated coils:
it occurs at $\lambda=0.5$, {\sl i.e.} the resonance frequency of a
single coil, but carries twice the weight. The left peak in the figure
corresponds to two horizontal inductances in series, whereas the right
peak corresponds to two parallel inductances separated by one lattice
spacing in the $x$-direction ({\sl i.e.} by capacities on either
side). From duality, we know that the resonances are the same in the
two cases but each time, only one peak contributes to the spectrum,
while the other resonance has zero cross section. These results can be
understood qualitatively in terms of, respectively, two inductances in
series (inductance $\simeq 2L$, with resonance pulsation
$\omega_{a}\simeq\omega_{0}/\sqrt{2}$), and two inductances in
parallel (inductance $\simeq {L}/{2}$, with resonance pulsation
$\omega_{b}\simeq{\omega_{0}}\cdot{\sqrt{2}}$), where
$\omega_{0}=1/\sqrt{L C}$ is the resonance pulsation of a single $LC$
circuit. This simple picture yields $\lambda_{a}=0.3692$ and
$\lambda_{b}=0.6410$, and almost fulfills the exact duality relation
$\lambda_{a}+\lambda_{b}=1$.  Note also that the maxima of the left
and right peaks are approximately the same, both being very close in
height to the central peak (which carries the intensity of two well
separated coils): despite the coupling, the amplitudes are almost
additive.

Fig.~\ref{fig3}b illustrates that the coupling between the bonds
decreases very quickly with the distance between the coils. The spectra
were obtained with similar coil-capacity configurations as the outer
peaks of Fig.~\ref{fig3}a, but each time the separation between the
coils was increased by interposing a capacity (empty link), shifting
the peaks closer to the central value of $\lambda=0.5$, characteristic
for independent coils.

Fig.~\ref{fig3}c represents the spectra of three different staircase
animals. The solid line corresponds to the spectrum of a single right
angle -- see animal 2b on Fig.~\ref{fig1}; for comparison, its
amplitude has been multiplied by a factor of eight. The dashed line
shows the resonances of an animal corresponding to number 3d on
Fig.~\ref{fig1} (the amplitudes are multiplied by a factor of
four). Finally, the dotted-dashed line represents the response of a
four-step staircase. Clearly, the relevant eigenvalues, {\sl i.e.}
those with non-vanishing cross sections, drift further and further
apart as the size of the animal increases, while the corresponding
cross-sections are roughly proportional to the number of links in the
$y$-direction. 
Besides the main resonances visible in the figure, the animals ---
especially the larger ones --- may contain eigenvalues of zero or
almost zero weight.
Note that all staircase animals are self-dual (if contained in an
infinite lattice); hence the eigenvalues occur in pairs satisfying
$\lambda_{1}+\lambda_{2}=1$, which provides a good check for the
numerical results and yields spectra symmetric about $\lambda=0.5$.

However, if such right-angle 2-animals are randomly spread on the
lattice the resulting impedance is quite different from an individual
animal's: the full line in Fig.~\ref{fig3}d corresponds to $36$
right-angle animals distributed at random on a $12\times 12$ network
(averaged over one hundred realizations) and only the peak near
$\lambda \simeq 0.7$ can be identified as a signature of the
right-angle structure. With the same conditions, but with $36$
horizontal bonds and $36$ vertical bonds chosen at random, the result
is given by the dashed line curve of Fig.~\ref{fig3}d.

We conclude that only in the very ``dilute'' limit ($p\ll 1$), the
spectral features may be assigned to elementary animals, such as
isolated links or pairs of links. If an animal is predominant in a
network, it is easy to predict the conductivity ratio $h$, and the
frequency which leads to the optimal conductivity: in practical
applications, a pattern could be chosen in order to obtain a selective
absorption or reflectivity for a given frequency band. This property
might have some interesting practical implications which could result
in the engineering of devices whose impedances might almost be chosen
{\sl \`a la carte}. As soon as the density of animals in the lattice
increases, and the animals' legs occupy in the order of $5\%$ of the
lattice bonds or more, the coupling between patterns increases and
with it the difficulty to assign the observed spectrum to a specific
animal.

We want to draw the reader's attention to the spectra that emerge if
the coils are not randomly distributed but regularly arranged on the
lattice, thus forming a regular super-structure of the dielectric
lattice.
\begin{figure}
\caption{(a) Real part of the impedance $Re(Z)$ versus
$\lambda$ for a $12\times 12$ lattice wholly occupied by capacities,
except for an array of $36$ regularly distributed horizontal coils,
linking sites $(x,y)$ and $(x,y+1)$ with $x$ and $y$ even.
$Re(Z)$ presents a ``colossal'' resonance for $\lambda\simeq 0.628$. 
(b) Same curves for an array where each horizontal coil originates at
a node $(x,y)$ with both coordinates $x$ and $y$ multiples of 3 (solid
line) and 4 (dotted line).
Figure (c) and (d) correspond, respectively, to Figure (a) and (b),
with right-angle animals replacing the horizontal coils.
The two peaks in Figure (c) occur at the same $\lambda$ as the
resonances of a single right-angle (see Fig.~\ref{fig3}c). These
resonances splits up when the periodicity of the coil array changes. 
}
\label{fig4}
\begin{center}
\epsfig{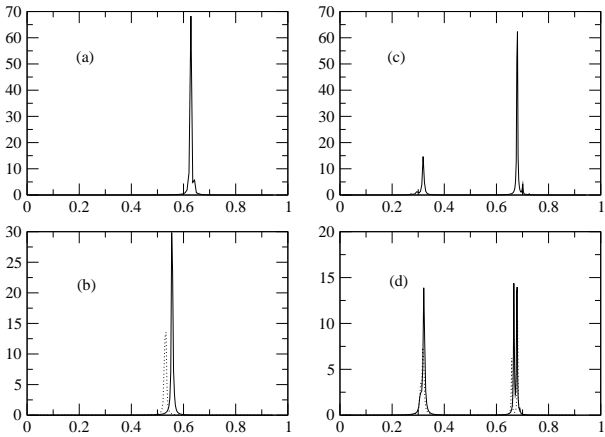}
\end{center}
\end{figure}
Fig.~\ref{fig4}a was obtained with an array of alternating bonds
(coils) and voids (capacities) on every other horizontal line of the
network, resulting in a density of inductive bonds of $12.5\%$ (every
second horizontal bond on every second horizontal line is occupied,
whereas all vertical bonds are empty). One naively expects a number of
resonances comparable to the number of coils, but this is not the
case: almost the entire spectral weight is concentrated in one single
peak occuring at $\lambda=0.628$. This result turns out to be robust
against variations of the lattice size: is has been confirmed by three
different arrays, $N_x=N_y=12, 18$ and $24$, and each time the domain
$\lambda\in[0,1]$ has been resolved into 500 intervals. The height of
the peak turns out to be almost invariant, since it is a function of
the density of coils contained in the array.  Note however that the
given values for the peak position are approximate, since it is
numerically rather difficult to locate this very narrow maximum
precisely.

In an analogous system, where in either direction the spacing between
the coils has been increased from 2 (bond-void) to 3 (bond-void-void)
or 4 (bond-void-void-void), very similar results are observed: the
spectra of these structures are shown in Fig.~\ref{fig4}b and present
peaks at, respectively, $\lambda_{\rm max}=0.556$ and $\lambda_{\rm
max}=0.533$. The slight shift to lower $\lambda$ may be easily
interpreted as a result of the decreasing inter-coil coupling when
more and more capacities are interposed.

In Fig.~\ref{fig4}c and d, the spectra for other regular arrangements
of bonds and voids are shown: Fig.~\ref{fig4}c is the result for an
analogous pattern as in Fig.~\ref{fig4}a, but with each bond replaced
by a right angle (see diagram 9 in Fig.~\ref{fig2}), thus resulting in
a periodic structure of right angles, separated in $x$ and
$y$-direction by a single void (capacity). At the first glance, the
results seem to be quite similar to those of the array of horizontal
bonds (Fig.~\ref{fig4}a), and instead of a single peak we observe two
peaks at $\lambda_{\rm max1}=0.318$ and $\lambda_{\rm max2}=0.668$.  A
closer examination shows that these values correspond to the
eigenvalues of a single ``right-angle animal'' alone.  For the same
structure, but with a horizontal and vertical periodicity of 3 and 4
(intercalation of, respectively, two or three capacities in either
direction), the spectra are shown in Fig.~\ref{fig4}d. We observe
that, in comparison to Fig.~\ref{fig4}c, both peaks split up slightly:
this illustrates the primordial role of coupling via interference in
the spectra of crystal-like arrangements of elementary lattice
animals.

One might summarize that the periodic arrangement of bonds and voids
could allow for the construction of composites with absorption
properties at very specific frequencies (see Fig.~\ref{fig4}a-d),
whereas disordered arrangements on the dielectric surface (as in
Fig~\ref{fig3}d) lead to multiple spectral peaks, which in the
infinite limit will turn into bands.

\section{Binary lattice of resistances and capacitors}
\label{sect3}

In this section, we present simulations for a disordered
resistance-capacitor system performed with the spectral method first
implemented in 2D by Jonckheere and Luck \cite{Luck} and then
enhanced to 3D by two of the authors \cite{Laurent}. This method
yields the total network impedance for any ratio $h$ of the local
impedances. Thereafter the Nyquist diagram can be readily drawn and
compared to experiments on a real sample. 
\begin{figure}
\caption{Nyquist diagrams, {\sl i.e.} a plot of $Im(Z(\omega))$
vs. $Re(Z(\omega))$ with $\omega$ as a parameter, for 2D and 3D
hypercubic resistor lattices, doped with various concentrations $p$
of capacitor bonds. }
\label{fig5}
\begin{center}
\epsfig{figure=figure5a_animaux.eps,width=7cm}
\end{center}
\begin{center}
\epsfig{figure=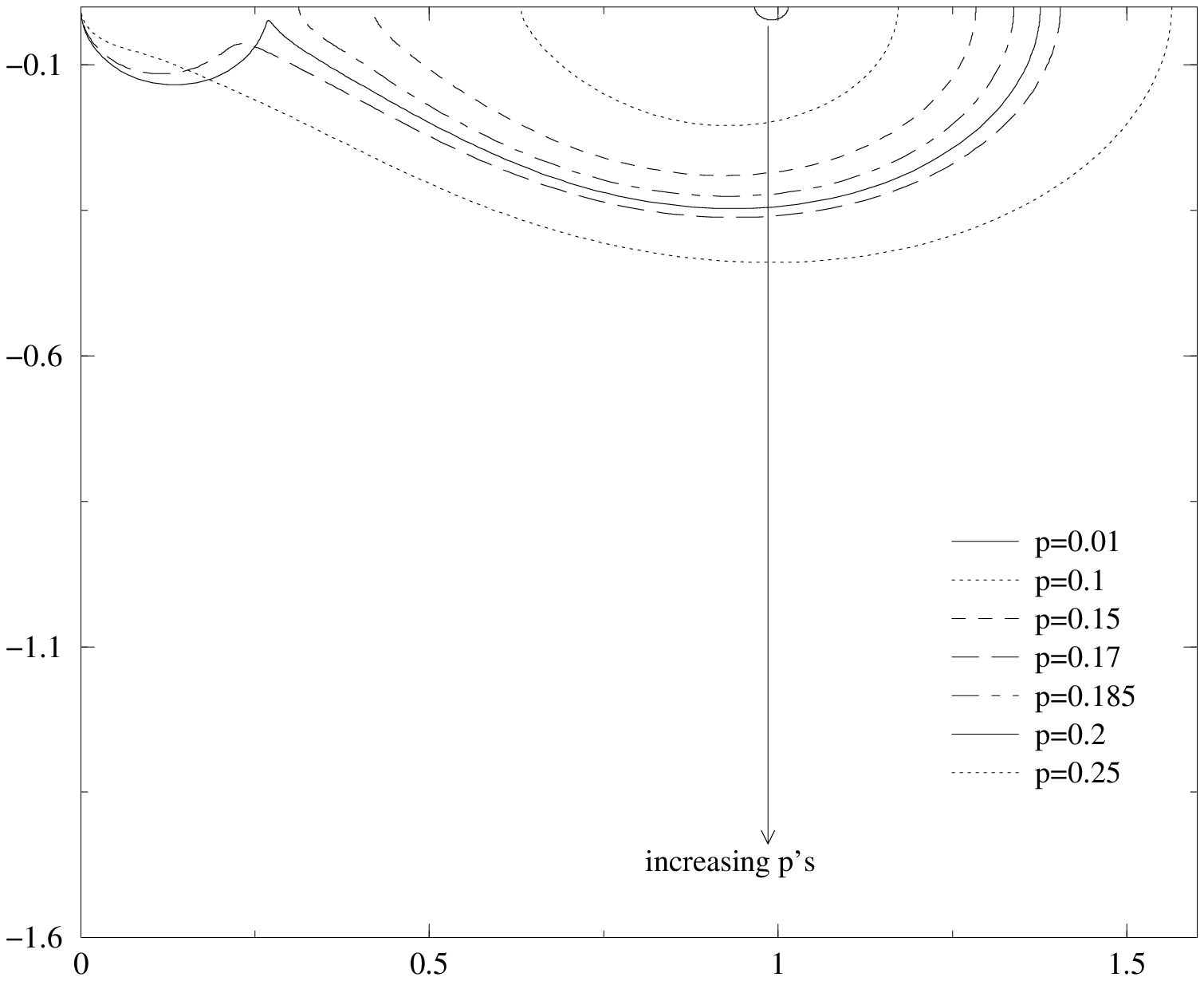,width=7cm}
\end{center}
\end{figure}
As a first application, Fig.~\ref{fig5} illustrates the Nyquist
diagrams obtained for a square and a cubic lattice of resistors
($\sigma_0=R^{-1}$) which are replaced (doped) with a probability
(concentration) $p$ by capacitors ($\sigma_1={\rm j}C\omega$).

The calculations are performed for $R=1$ and $C=1$ and lattice sizes
of $12\times12$ in 2D and $8\times8\times8$ in 3D. Since individual
calculations for a random distribution are meaningless, averages are
taken over a thousand samples for each density $p$, with $p$ varying
from low density $p=0.01$ to the percolation threshold $p\simeq p_{c}$
(with the thresholds $p_{c}=0.5$ in 2D and $\simeq 0.25$ in 3D up to
finite size corrections). 

Fig.~\ref{fig5} shows that for a very small amount of capacitors
($p=0.01$), the Nyquist diagram is a perfect semi-circle centered on
the value $R$ corresponding to the DC resistivity of the sample. 
\begin{figure}
\caption{Simple electronic representations and their schematic Nyquist
diagrams. The upper figure represents a 2D network where the
conductors percolate. The second figure corresponds to the same
situation when the insulators percolate.  The last one is a model for
a 3D system, with probability p such that conductors and insulators
simultaneously percolate.}
\label{fig6}
\begin{center}
\epsfig{figure=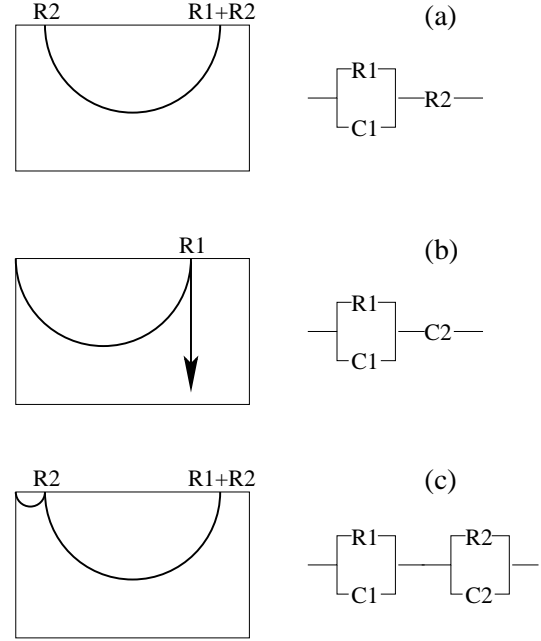,width=7cm}
\end{center}
\end{figure}
This can be interpreted in terms of a single capacity in parallel to a
resistance, a system which is governed by a single time constant
$\tau=RC$ (see Fig.~\ref{fig6}a). For increasing $p$, the radius of
the semi-circle becomes larger. As the percolation threshold is
approached (at which the two electrodes would be connected by a
percolating path of capacitors), a distortion of the semi-circle
becomes perceptible at high frequency ({\sl i.e.} close to the
origin). This can be understood since the impedance of a single
capacitor vanishes at infinite frequency (see Fig.~\ref{fig6}b).

In 2D, where necessarily either capacities or resistances percolate
(except for finite size effects which are compensated by averaging
over many samples), all Nyquist diagrams can be interpreted
qualitatively by the phenomenological model of Fig.~\ref{fig6}.
If the resistances percolate (see Fig.~\ref{fig6}a), the impedance
corresponds at low frequency to $R_1$ and $R_2$ in series, thus
equaling $R_1+R_2$, whereas at high frequency the impedance is given
by $R_2$ alone since $C_1$ has zero impedance.
By contrast, if the capacities percolate, the equivalent
phenomenological circuit is shown in Fig.~\ref{fig6}b. At high
frequency, the circuit's impedance is governed by the two capacities
in series and thus zero; at low frequency, by contrast, the real part
is given by $R_1$ while the capacities insulating behaviour gives rise
to an infinite imaginary part.

In 3D, the behaviour is radically different from the 2D case: a second
semi-circle ending at the origin $Z=0$ appears in Fig.~\ref{fig5}b as
the percolation threshold $p_c=0.2488126$ \cite{Lore} is approached
from below. It is a signature of growing clusters in which capacities
percolate, but which are still embedded in a resistor-dominated
environment (see Fig.~\ref{fig6}c for an illustration of such a
cluster). The ratio of the radii of the two semi-circles is related to
the percolation probability for a finite size sample.  
Above $p_{c}$ both species may percolate simultaneously. The resulting
Nyquist diagram contains a single structure which is albeit more
complicated than a simple semi-circle, since it is determined by a
whole set of time constants deforming the diagram close to the origin
({\sl i.e.} at high frequency).

\section{Conclusion}

As shown in Ref.~\cite{Vinograd,Sarych}, the exact numerical
renormalization (ENR) method allows for the calculation of the
impedances of disordered networks, and is particularly well adapted if
one species dominates the network. In the present work, we argue that
this method is very competitive, since the formalism is simple and
easy to implement, and remains numerically efficient even far from
its privileged domain of application. 
The spectral method, on the other hand, presents advantages for
repeated calculations with a great number of conductance ratios, since
the spectral density of a given sample can be stored in memory for
calculations of the conductances for several values of $h$ (or
$\lambda$). 

For the investigation of frequency-dependent quantities, these two
algorithms are clearly more suitable than classical approaches such as
effective medium approximation, sparse matrix method \cite{Albi},
transfer matrix method \cite{Norm} or the star-triangle transformation
\cite{Lobb}. The correctness of the ENR algorithm has been verified by
comparison to small clusters (animals) whose properties can be
calculated analytically.

In the present work, the ENR algorithm was used to calculate the
Nyquist diagrams of two and three dimensional hypercubic lattices of
randomly distributed resistors and capacities. Obviously, the results
are very sensitive to the proportion of capacities to resistors,
especially close to the percolation threshold. But also the
dimensionality of the system plays an important role: in 2D, the
Nyquist diagrams is essentially given by a single semi-circle; in 3D,
by contrast, for densities just below the percolation threshold, two
semi-circles are found which indicates the imminent simultaneous
percolation of both components, capacities and resistors, through the
system. In 2D, this scenario is ruled out, since on  sufficiently large
lattices, only one component may percolate at a time.

The Exact Numerical Renormalization proposed in
Ref. \cite{Vinograd,Sarych,Tort} is undoubtedly the most efficient
numerical method if bonds or voids form large clusters, as in the
vicinity of a percolation threshold. The algorithm also allows for a
detailed analysis of the inter-bond coupling which is particularly
efficient between linked bonds, but remains important for nearest
neighbor bonds. Complementary information is provided by the spectral
method, even though the ``global animal'' is in general too large for
its spectrum to be interpreted in simple terms \cite{Luck,Laurent}.
However, as soon as two conducting animals are separated by more than
one capacity, the coupling between the animals decreases rapidly, and
even if the influence of the coupling on the resonance positions is
still considerable, the total spectral weight is roughly given by the
sum of the individual animals' contributions.

Further complications arise from the presence of the electrodes: the
animal's spectrum depends on its distances from the electrodes (see
Tab.~\ref{tab2} and Tab.~\ref{tab3}), and the resonances may be
slightly altered if the animal is very close to the electrodes. For a
simple animal (constituted by a few legs) and localized near the
center of a large lattice (for instance $N_x=N_y=24$) the difference
between the results of the ENR algorithm (where the electrodes are
taken into account) and the exact analytical eigenfrequencies and
their weights are almost negligible, since the shift of the resonance
pulsations behaves as $1/N^2$.

As shown in Sect.~\ref{sect3}, for periodic arrangements of given
elementary animals, e.g. linked inductances in a capacitive array, the
spectra are dominated by a few colossal resonances only. These appear
at $\lambda$-values mirroring the periodicity of the arrangement and
the structure of the animal itself. Only for very special
arrangements, or in the very dilute limit, the whole spectrum may be
completely understood on the basis of the individual animals' spectra
themselves. As these super-lattices present very interesting spectral
properties, devices with specific absorption or transmission
properties for electromagnetic waves might be engineered as regular
arrays of metallic grains of a well-defined form on a dielectric
support. In the dilute limit $p\ll 1$, the devices properties would be
essentially given by the individual animals behaviour.

One can hope for very useful applications of these systems, especially
in nano-technology. Moreover, progress in furtivity, active skins,
giant Raman scattering could be spin-offs of a good understanding of
the behavior of composite materials in an electromagnetic field (see
Ref.~\cite{Brou}).  The knowledge of, for instance, the resonances of
given patterns would allow to construct skins absorbing incident
electromagnetic waves of well-defined wavelengths. By combining
several animals, one could obtain an arbitrary set of resonances,
leading to an arbitrary frequency response of the painted object.

\begin{acknowledgement}
Acknowledgements: It is a real pleasure to thank J.-M.~Luck,
A.K.~Sarychev and J.P.~Clerc for very useful discussions.
\end{acknowledgement}

\appendix
\section{Appendix: Construction of the animals}
In the following, we will present a recursive algorithm which
generates all $n+1$-animals based on the entire zoo of $n$-legged
animals. The basic idea is simple: it consists of adding a leg (or
bond) to each $n$-animal in every possible location. However, the zoo
of $n+1$-animals obtained in this manner will generally contain many
identical animals. These will be eliminated by comparing each new
$n+1$-animal to the already created ones, such that only one animal
per species is kept.  In the following, the method is applied to a 2D
square lattice; a generalization to higher dimensions or different
lattice types is however straightforward.

As recurrency seed, we use the only animal with ``zero'' legs, {\sl
i.e.}  a point at the origin. The zoo of 1-legged animals is then
obtained by adding a bond to this point, and contains two horizontal
and two vertical 1-animals. 

In an analogous way, the zoo of 2-legged animals can be generated by
adding a further bond to these four 1-animals. Since each 1-animal has
two endpoints and each of these endpoints has three vacancies at which
a bond can be added, there are $4\times 2\times 3=24$ 2-animals. A
closer look shows however that all 2-animals having their midpoint at
the origin ({\sl i.e.} 4 of the L-shaped and 2 of the straight ones)
are created twice by this procedure: the full zoo of 2-legged animals
thus contains only 18 species.
The number of species as a function of the number of legs resulting
from this recursive algorithm is listed in the third column in
Table~\ref{tabA1}.
\begin{table}
\caption{Number of animals and species in the $n$-legged zoo as a
function of $n$ and symmetry. For details, see text.}
\label{tabA1}
\begin{center}
\end{center}
\begin{tabular}{|c|r|r|r|r|r|}
\hline
     &                  & \multicolumn{4}{|c|}{species in the zoo} \\
$n$  &   animals        & \multicolumn{4}{|c|}{symmetries:}             \\
     &                  &  without &   +transl. &    +rot.  &   +mirror \\
\hline
1    &                4 &       4  &          2 &         1 &         1 \\
2    &               24 &      18  &          6 &         2 &         2 \\
3    &              192 &      88  &         22 &         7 &         5 \\
4    &           1\,920 &     439  &         88 &        25 &        16 \\
5    &          22\,784 &  2\,224  &        372 &        99 &        55 \\
6    &         311\,296 & 11\,342  &     1\,628 &       416 &       222 \\
7    &      4\,796\,416 & 58\,168  &     7\,312 &    1\,854 &       950 \\
8    &     82\,049\,024 &          &            &    8\,411 &    4\,265 \\
9    & 1\,539\,876\,352 &          &            &           &   19\,591 \\
\hline
\end{tabular}
\end{table}

By considering the system without electrodes, three more symmetries
are gained which allow us to reduce the number of species in each zoo.
(i) Translational invariance: in a zoo with electrodes, many animals
are connected to each other by a simple shift, e.g. the two 2-animals
which result from adding a horizontal bond either to the left or to
the right site of the horizontal 1-animal. In an infinite lattice,
such animals belong to the same species. To avoid double counting in
the generation procedure, each newly created $n+1$-animal will be
shifted to a well-defined place -- e.g. such that its lower left
corner has lattice coordinates $(0,0)$ -- and will then be compared to
the animals already contained in the zoo. As shown by the values in
column 4 of Table~\ref{tabA1}, translational invariance reduces the
number of $n$-legged species by roughly a factor $n+1$. This factor
corresponds to the number of sites occupied by an $n$-animal without
loops, and is thus exact for $n\le 3$ since a loop requires at least
four bonds.

(ii) Rotational invariance: by eliminating animals connected to each
other via rotations by multiples of $\pi/2$, the zoos can be further
reduced, as can be seen from column 5 in Table~\ref{tabA1}. The
reduction with respect to the previous column approaches a factor of
$4$ in the limit of large animals, $n\gg 1$. This can be easily
understood because a general animal can be placed in 4 different
orientations in a square lattice. This bare factor of $4$ is reduced
by the presence of animals which are self-invariant under rotations by
multiples of $\pi/2$. However, since the relative abundance of these
self-symmetric species in the zoo decreases as the number of legs $n$
increases, the reduction factor approaches $4$ for $n\to\infty$.

(iii) Mirror symmetry: still more species can be eliminated by
mirroring on the $x$-axis. As there is only one mirror axis, the
number of species in the last column of Table~\ref{tabA1} is reduced
with respect to the previous column by roughly a factor of $2$ (note
that mirroring on the $y$-axis is equivalent to mirroring on the
$x$-axis and a subsequent rotation by $\pi$). Deviations from this
factor of $2$ are again due to self-symmetric species, and diminish as
the size of the animals increases.

The last column in Table~\ref{tabA1} thus shows the number of
topologically distinct species in the zoo of $n$-legged animals. These
species along with the relative number of animals they contain are
shown, for $n\le 4$, in Fig.~\ref{fig1}.

\end{document}